\documentclass[twocolumn,runningheads]{svjour2}
\smartqed  
\usepackage{graphicx}
\def\be{\begin{equation}}
\def\ee{\end{equation}}
\def\bB{\mathbf B}

\def\RNS{R_{\rm NS}}

\def\simlt{\lower.5ex\hbox{$\; \buildrel < \over \sim \;$}}
\def\simgt{\lower.5ex\hbox{$\; \buildrel > \over \sim \;$}}

\def\Ldiss{L_{\rm diss}}

\def\>{$>$}
\def\<{$<$}

\def\simlt{\lower.5ex\hbox{$\; \buildrel < \over \sim \;$}}
\def\simgt{\lower.5ex\hbox{$\; \buildrel > \over \sim \;$}}
\def\ch2{$\chi^{2}$}

\def\ee{\'{e}}

\def\sT{\sigma_{\rm T}}
\def\dd{{\rm d}}
                                                                                
\def\be{\begin{equation}}
\def\ee{\end{equation}}

\def\bB{{\,\mathbf B}}
\def\bE{{\,\mathbf E}}
\def\bj{{\,\mathbf j}}

\def\jB{j_B}
\def\RNS{R_{\rm NS}}
\def\gres{\gamma_{\rm res}}

\def\Rmax{R_{\rm max}}

\def\lres{l_{\rm res}}
\def\Lres{a_{\rm res}}

\def\lD{\lambda_{\rm D}}

\def\lCoul{\ell_{\rm Coul}}

\def\kB{k_{\rm B}}

\def\tdyn{t_{\rm dyn}}
                                                                                
\newbox\grsign \setbox\grsign=\hbox{$>$} \newdimen\grdimen \grdimen=\ht\grsign
\newbox\simlessbox \newbox\simgreatbox \newbox\simpropbox
\setbox\simgreatbox=\hbox{\raise.5ex\hbox{$>$}\llap
     {\lower.5ex\hbox{$\sim$}}}\ht1=\grdimen\dp1=0pt
\setbox\simlessbox=\hbox{\raise.5ex\hbox{$<$}\llap
     {\lower.5ex\hbox{$\sim$}}}\ht2=\grdimen\dp2=0pt
\setbox\simpropbox=\hbox{\raise.5ex\hbox{$\propto$}\llap
     {\lower.5ex\hbox{$\sim$}}}\ht2=\grdimen\dp2=0pt
\def\simgt{\mathrel{\copy\simgreatbox}}
\def\simlt{\mathrel{\copy\simlessbox}}

\begin{document}

\title{Magnetar Corona}



\author{A. M. Beloborodov \and C. Thompson}


\institute{A. M. Beloborodov\at 
              Physics Department and Columbia Astrophysics Laboratory,
              Columbia University, 538 W 120th Street, New York, US \\
              \email{amb@phys.columbia.edu}  
              \and 
              C. Thompson\at
              Canadian Institute for Theoretical Astrophysics,
              University of Toronto, 60 St. George Street, Toronto,
              Canada; 
              \email{thompson@cita.utoronto.ca}
}

\date{Received: date / Accepted: date}

\maketitle

\begin{abstract}
  Persistent high-energy emission of magnetars is produced by a plasma corona 
  around the neutron star, with total energy output of $\sim 10^{36}$~erg/s. 
  The corona forms as a result of occasional starquakes 
  that twist the external magnetic field of the star and induce electric 
currents in the closed magnetosphere. Once twisted, the magnetosphere
cannot untwist immediately because of its self-induction. The self-induction
electric field lifts particles from the stellar surface, accelerates
them, and initiates avalanches of pair creation in the magnetosphere.
The created plasma corona maintains the electric current demanded by
${\rm curl}\bB$ and regulates the self-induction e.m.f. by screening.
This corona persists in dynamic equilibrium: it is continually lost to
the stellar surface on the light-crossing time $\sim 10^{-4}$~s and
replenished with new particles. In essence, the twisted magnetosphere
acts as an accelerator that converts the toroidal field energy to
particle kinetic energy. 
The voltage along the magnetic field lines is maintained near threshold for 
ignition of pair production, in the regime of self-organized criticality.
The voltage is found to be about $\sim 1$~GeV
which is in agreement with the observed dissipation rate $\sim 10^{36}$~erg/s. 
The coronal particles impact the solid
crust, knock out protons, and regulate the column density of the
hydrostatic atmosphere of the star. The transition layer between the
atmosphere and the corona 
is the likely source of the observed 100~keV emission
from magnetars. 
The corona also emits curvature radiation up to $10^{14}$~Hz
and can supply the observed IR-optical luminosity. 
\end{abstract}

\section{Introduction}
\label{intro}

At least 10\% of all neutron stars are born as magnetars, with
magnetic fields $B>10^{14}$~G.
Their activity is powered by the decay of the
ultrastrong field and lasts about $10^4$~years. They are observed at this
active stage as either Soft Gamma Repeaters (SGRs) or Anomalous X-ray
Pulsars (AXPs) \cite{WT06}.

Besides the sporadic X-ray outbursts, a second, persistent, form of
activity has been discovered by studying the emission spectra of
magnetars. Until recently, the spectrum was known to have a thermal
component with temperature $\kB T\sim 0.5$~keV, interpreted as blackbody
emission from the star's surface. The soft X-ray spectrum also showed a
tail at $2-10$~keV with photon index $\Gamma=2-4$. 
This deviation from
pure surface emission already suggested that energy is partially released
above the star's surface.
Recently, observations by RXTE and INTEGRAL have revealed even more
intriguing feature: magnetars are bright, persistent sources of 100 keV
X-rays \cite{Kuiper06}.
This high-energy emission forms a separate component of the magnetar spectrum.
It becomes dominant above 10~keV,
has a very hard spectrum, $\Gamma\simeq 1$, and peaks above 100~keV
where the spectrum is unknown.
Its luminosity, $L\sim 10^{36}$~erg~s$^{-1}$, even exceeds
the thermal luminosity from the star's surface. The observed hard X-rays
can be emitted only in the exterior of the star and demonstrate
the presence of an active plasma corona.
                                                                                
This corona can form as a result of starquakes that shear the magnetar
crust and its external magnetic field.
The persistence of magnetospheric twists can explain several observed
properties of magnetars.
Thompson, Lyutikov, \& Kulkarni~\cite{TLK} investigated the observational 
consequences using a static, force-free model, idealizing the magnetosphere 
as a globally twisted dipole. They showed that a twist affects the spindown 
rate of the neutron star: it causes the magnetosphere to flare out slightly 
from a pure dipole configuration, thereby increasing the braking torque 
acting on the star.
Although the calculations of force-free configurations are independent of 
the plasma behavior in the magnetosphere, they rely on the presence of 
plasma that can conduct the required current $\bj_B=(c/4\pi)\nabla\times\bB$.
This requires a minimum particle density $n_c=j_B/ec$. 
Thompson et al.~\cite{TLK} showed that even this minimum density can 
modify the stellar radiation by multiple resonant scattering.
                                                                                
The problem of plasma dynamics in the closed twisted magnetosphere has been
formulated in \cite{BT}, and a simple solution has been found to this problem.
It allows one to understand the observed 
energy output of the magnetar corona and the evolution of magnetic twists. 
It is worth mentioning that plasma behavior around neutron stars has been 
studied for decades in the context of radio pulsars, and that problem 
remains unsolved. The principle difference with radio pulsars is that
their activity is caused by {\it rotation}, and
dissipation occurs on {\it open magnetic lines} that connect the star
to its light cylinder. Electron-positron pairs are then created on open
field lines \cite{RS75,AS79}.
By contrast, the formation of a corona around
a magnetar does not depend on its rotation. All observed
magnetars are slow rotators ($\Omega=2\pi/P\sim 1$~Hz), and their
rotational energy is unable to feed the observed coronal emission.
Practically the entire plasma corona is immersed in the {\it closed} 
magnetosphere and its heating must be caused by some form of dissipation 
on the closed field lines. It is this closure 
that facilitates the corona problem by providing both boundary conditions 
at the two footpoints of 
 a field line.
                                                                                
The basic questions that one would like to answer are as follows.
How is the magnetosphere populated with plasma? The neutron star
surface has a temperature $k_{\rm B}T\simlt 1$~keV and the scale-height
of its atmospheric layer (if it exists)
is only a few cm.  How is the plasma supplied
above this layer and what type of particles populate the corona?
How is the corona heated, and what are the typical energies of the particles?
If the corona conducts a current associated with a magnetic
twist $\nabla\times\bB\neq 0$, how rapid is the dissipation of this
current, i.e. what is the lifetime of the twist? Does its decay imply the
disappearance of the corona?
                                                                                
An outline of the proposed model is as follows.
The key agent in corona formation is an electric field $E_\parallel$
parallel to the magnetic field. $E_\parallel$ is generated by the
self-induction of the gradually decaying current and in essence
measures the rate of the decay. It determines the heating rate of
the corona via Joule dissipation. If $E_\parallel=0$ then the corona is
not heated and, being in contact with the cool stellar surface, it will
have to condense to a thin surface layer with $k_{\rm B}T\simlt 1$~keV.
The current-carrying particles cannot flow upward against
gravity unless a force $eE_\parallel$ drives it. On the other hand, when
$E_\parallel$ exceeds a critical value, $e^\pm$ avalanches
are triggered in the magnetosphere, and the created pairs screen the electric
field. This leads to a ``bottleneck'' for the decay of a magnetic twist,
which implies a slow decay.
                                                                                
Maintenance of the corona and the slow decay of the magnetic twist
are intimately related because both are governed by $E_\parallel$.
In order to find $E_\parallel$, one can use Gauss' law
$\nabla\cdot\bE=4\pi\rho$ where $\rho$ is the net charge density of
the coronal plasma. This constraint implies that
$\bE$ and $\rho$ must be found self-consistently. The problem turns out to be
similar to the classical double-layer problem of plasma physics
with a new essential ingredient: $e^\pm$ creation.
                                                                                
A direct numerical experiment can be designed that simulates 
the creation and behavior of the plasma in the magnetosphere. 
The experiment shows how the plasma and
electric field self-organize to maintain the time-average magnetospheric
current $\bar{\bj}=\bj_B$ demanded by the magnetic field,
$(4\pi/c)\bar{\bj}=\nabla\times\bB$. The electric current admits no steady
state on short timescales and keeps fluctuating, producing $e^\pm$ avalanches.
This state may be described as a self-organized criticality.
Pair creation is found to provide a robust mechanism for limiting the voltage
along the magnetic lines to $e\Phi_e\simlt 1$~GeV
and regulate the observed luminosity of the corona.

\section{Mechanism of corona formation}

A tightly wound-up magnetic field is assumed to exist inside magnetars
\cite{TD01}.
The internal toroidal field can be
much stronger than the external large-scale dipole
component.
The essence of magnetar activity is the transfer of magnetic helicity
from the interior of the star to its exterior, where it dissipates.
This involves rotational motions of the crust, which inevitably twist
the external magnetosphere anchored to the stellar surface.
The magnetosphere is probably twisted in a catastrophic way, when the
internal magnetic field breaks the crust and rotates it. Such starquakes
are associated with observed X-ray bursts \cite{TD95}.
The most interesting effect of a starquake for us here is the partial
release of the winding of the internal magnetic field into the exterior,
i.e., the injection of magnetic helicity into the magnetosphere.
                                                                                
Since the magnetic field is energetically dominant in the magnetosphere,
it must relax there to a force-free configuration with $\bj\times\bB=0$.
Electric current $\bj$ initiated by a starquake flows
along the magnetic field lines to the exterior of the star, reaches the top 
of the field line, and comes back to the star at the other footpoint, 
then enters the atmosphere and the crust,
following the magnetic field lines ($\bj\parallel\bB$).
Sufficiently deep in the crust the current flows across the field lines
and changes direction so that it can flow back to the twisted footpoint
through the magnetosphere (or directly through the star, depending on the
geometry of the twist).
The current can flow across the field lines in the deep crust because it 
is able to sustain a significant Amp\`ere force $\bj\times\bB/c\neq 0$ 
--- this force is balanced by elastic forces of the deformed crust. 

The emerging currents are easily maintained during the X-ray outburst
accompanying a starquake. A dense, thermalized plasma is then present in
the magnetosphere, which easily conducts the current. Plasma remains
suspended for some time after the starquake because of the transient
thermal afterglow \cite{I01}.
Eventually the afterglow extinguishes and the radiative flux becomes unable
to support the plasma outside the star. The decreasing density then threats
the capability of the magnetosphere to conduct the current of the
magnetic twist. A minimal ``corotation'' charge density
$\rho_{\rm co} = -{\bf\Omega}\cdot\bB/2\pi c$
is always maintained \cite{GJ69},
but it is far smaller than
needed to supply the current 
$\bj\sim (c/4\pi)(B/\RNS)$. Can the lack of charges stop the flow of
electric current? A simple estimate shows that the current cannot stop
quickly because of its self-induction. 
A slow decay of the current generates a sufficient self-induction voltage 
that helps the magnetosphere to re-generate the plasma that carries the 
current, so that the twisted force-free configuration 
persists with a qausi-steady $\bj\neq 0$.
                                                                                
The stored energy of non-potential (toroidal) magnetic field associated with
the ejected current is subject to gradual dissipation.
In our model, this dissipation
feeds the observed activity of the corona.
The stored energy is concentrated near the star and carried
by closed magnetic lines with a maximum extension radius
$R_{\max}\sim 2R_{NS}$.
Thus, most of the twist energy will be released if these lines
untwist, and we focus here on the near magnetosphere
$R_{\max}\sim 2R_{NS}$.
                                                                                
Consider a magnetic flux tube with cross section
$S\simlt \RNS^2$ and length $L\sim\RNS$ which carries a current $I=Sj$.
The stored magnetic energy of the current per unit length of the tube is
\be
   {{\cal E}_{twist}\over L} \sim \frac{I^2}{c^2}\, S.
\ee
The decay of this energy is associated with an electric field parallel to
the magnetic lines $\bE_\parallel$: this field can accelerate particles
and convert the magnetic energy into plasma energy. Conservation of energy
can be expressed as
\be
  \frac{\partial}{\partial t}\left(\frac{B^2}{8\pi}\right)=-\bE\cdot\bj
         -\nabla\cdot\left(c\frac{\mathbf E\times \mathbf B}{4\pi}\right),
\ee
as follows from Maxwell's equations with $E\ll B$.
The first term on the right-hand side is the Joule dissipation caused by
$\bE_\parallel$. The second term --- the divergence of the Poynting flux --- 
vanishes for the induced field $\bE_{\parallel}$.

\begin{figure}
\centering
\includegraphics[width=0.48\textwidth]{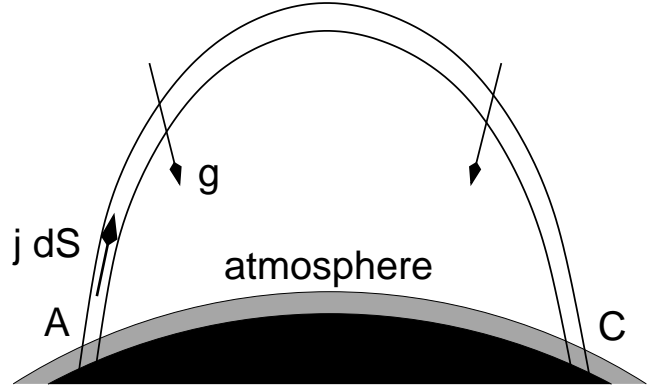}
\caption{
Schematic picture of a current-carrying magnetic tube 
anchored to the star's surface.
The current is initiated by a starquake that twists one (or both)
footpoints of the tube. The current flows along the tube up to the
magnetosphere and comes back to the star at the other footpoint.
A self-induction voltage is created along the tube
between its footpoints, which accompanies the gradual decay of the current.
The voltage is generated because the current has a tendency to become
charge-starved above the atmospheric layer whose scale-height
$h=k_{\rm B}T/gm_p$ is a few cm.}
\label{fig:1}       
\end{figure}

The decay of the twist is related to the
voltage between the footpoints of a magnetic line,
\be
\label{eq:Phi_e}
   \Phi_e=\int_A^C {\mathbf E}\cdot \dd{\mathbf l}.
\ee
Here A and C are the anode and cathode footpoints and $\dd{\mathbf l}$
is the line element; the integral is taken along the magnetic line outside
the star (Fig.~1).  The product $\Phi_e I$ approximately represents
the dissipation rate $\dot{\cal E}$ in the tube.  (The two quantities are
not exactly equal in a time-dependent circuit.)
The instantaneous dissipation rate is given by
\be
  \dot{\cal E}=\int_A^C {\mathbf E}\cdot {\mathbf I}(l) \dd{\mathbf l},
\ee
where ${\mathbf I}(l)=S{\mathbf j}$ is the instantaneous current at
position $l$ in the tube. The current is fixed at footpoints A and C, 
$I=I_0$; it may, however, fluctuate between the footpoints.
                                                                                
The voltage $\Phi_e$ is entirely maintained by the self-induction that
accompanies the gradual untwisting of the magnetic field. Thus, $\Phi_e$ 
reflects the gradual decrease of the magnetic helicity in the tube,
\be
\dot{\cal H} = {\partial\over\partial t}\int_A^C {\bf A}\cdot{\bf B}\, \dd V
= -\int_A^C {\bf E\cdot B}\, S(l)\, \dd l.
\ee
(Here $\dd V = S(l)\dd l$ is the volume element.)  It also
determines the effective resistivity of the tube: ${\cal R}=\Phi_e/I$.
The self-induction voltage passes the released magnetic energy to charged
particles, and a higher rate of untwisting implies a higher energy $e\Phi_e$
gained per particle. A huge magnetic energy is stored in the twisted tube,
and a quick untwisting would lead to extremely high Lorentz factors of the
accelerated particles.
                                                                                
There is, however, a bottleneck that prevents a fast decay of the twist:
the tube responds to high voltages through the copious production of
$e^\pm$ pairs.  Runaway pair creation is ignited when electrons are
accelerated to a certain Lorentz factor $\gamma_\pm\sim 10^3$
(see below).
The created $e^\pm$ plasma does not tolerate large $E_\parallel$ --- the
plasma is immediately polarized, screening the electric field.
This provides a negative feedback that limits the magnitude of $\Phi_e$
and buffers the decay of the twist.
                                                                                
A minimum $\Phi_e$ is needed for the formation of a corona.
Two mechanisms can supply plasma: (1) Ions and electrons 
can be lifted into the magnetosphere from the stellar surface. This 
requires a minimum voltage,
\be
  e\Phi_e \sim gm_i R_{\rm NS} \sim 200\;\;{\rm MeV},
\ee
corresponding to $E_\parallel$
that is strong enough to lift ions (of mass $m_i$) from the anode 
footpoint and electrons from the cathode footpoint.
(2) Pairs can be created in the magnetosphere if
\be
  e\Phi_e \sim \gamma_\pm mc^2 = 0.5\gamma_\pm\;\;{\rm MeV},
\ee
which can accelerate particles to the
Lorentz factor $\gamma_\pm$ sufficient to ignite $e^\pm$ creation.
                                                                                
If $\Phi_e$ is too low and the plasma is not supplied, a flux tube is
guaranteed to generate a stronger electric field. The current becomes
slightly charge-starved: that is, the net density of free charges becomes
smaller than $|\nabla\times\bB|/4\pi e$.
The ultrastrong magnetic field, whose twist carries an enormous energy
compared with the energy of the plasma, does not change and responds
to the decrease of $\bj$ by generating an electric field: a small
reduction of the conduction current  $\bj$ induces a displacement current
$(1/4\pi)\partial {\mathbf E}/\partial t = (c/4\pi)\nabla\times\bB -\bj$.
The longitudinal electric field
$E_\parallel$ then quickly grows until it can pull
particles away from the stellar surface and ignite pair creation.
                                                                                
The limiting cases $E_\parallel\rightarrow 0$ (no decay of the twist)
and $E_\parallel\rightarrow\infty$ (fast decay) both imply a contradiction.
The electric field and the plasma content of the corona must regulate each
other toward a self-consistent state, and the gradual decay of the twist
proceeds through a delicate balance: $E_\parallel$ must be strong enough
to supply plasma and maintain the current in the corona;
however, if plasma is oversupplied $E_\parallel$ will be reduced by
screening.  A cyclic behavior is possible, in which plasma is periodically
oversupplied and $E_\parallel$ is screened. Our numerical experiment
shows that such a cyclic behavior indeed takes place.

\section{Electric circuit: numerical experiment}

Three facts facilitate the simulation of the 
coronal electric circuit:
                                                                                
1. The ultrastrong magnetic field makes the particle dynamics 1-D.
The magnetic field lines are not perturbed significantly
by the plasma inertia, and they can be thought of as fixed ``rails'' along
which the particles move. The particle motion is confined to the
lowest Landau level and is strictly parallel to the field (the lifetime
of a particle in an excited Landau state is tiny).
                                                                                
2. The particle motion is collisionless in the magnetosphere.
It is governed by two forces only: the component of gravity projected onto
the magnetic field and a collective electric field $E_\parallel$ which is
determined by the charge density distribution and must be found
self-consistently.
                                                                                
3. The star possesses a dense and thin atmospheric layer.\footnote{Such a
layer initially exists on the surface of a young neutron star. Its
maintenance is discussed in \cite{BT}.}
Near the base of the atmosphere, the required current
$\bj_B=(c/4\pi)\nabla\times {\mathbf B}$ is easily conducted,
with almost no electric field. Therefore the circuit
has simple boundary
conditions: $E_\parallel=0$ and fixed current. The atmosphere
is much thicker than the skin depth of the plasma and
screens the magnetospheric electric field from the star.
                                                                                
Our goal is to understand the plasma behavior above the screening
layer, where the atmospheric density is exponentially reduced and
an electric field $E_\parallel$ must develop.
The induced electric field $\bE$ and conduction current $\bj$ satisfy the
Maxwell equation,
\be
\label{eq:Maxw_}
  \nabla\times \bB=\frac{4\pi}{c}\bj
                   +\frac{1}{c}\frac{\partial \bE}{\partial t}.
\ee
Here $\bj$ is parallel to the direction of the magnetic field, and
the force-free condition requires $\nabla\times \bB\parallel\bB$.
Therefore,
in the fixed magnetic configuration where $\nabla\times \bB$ does not vary
with time, $\partial\bE/\partial t$ is parallel to $\bB$, i.e.
only $E_\parallel$ is created by the self-iduction effect.
We consider a simple model problem where no pre-existing
perpendicular field $\bE_\perp$ exists; more precisely, we require
$\nabla\cdot\bE_\perp=0$. Then Gauss' law reads
\be
\label{eq:Gauss1}
   4\pi\rho=\nabla\cdot\bE_\parallel=\frac{\dd E_\parallel}{\dd l},
\ee
where $l$ is length measured along the magnetic tube. 
Then the problem becomes strictly 1-D since $\bE_\perp$ has no
relation to charge density and falls out from the problem.
Note that the approximation of 1D circuit, $\jB(t)=const$, where
$\jB\equiv (c/4\pi)|\nabla\times\bB|$,
excludes the excitation of transverse waves in the magnetosphere,
which in reality can exist.
These waves are described by the coupled fluctuations $\delta\jB$
and $\bE_\perp$ on scales much smaller than the circuit size $L$.
(Wavelengths $\lambda\sim c/\omega_P$ can be excited by
plasma oscillations.)  These fluctuations may be expected to have a
small effect on the circuit solution
unless $\delta\jB$ becomes comparable to $\jB$.
                                                                                
With $\jB(t)=const$, the particle motions
on different magnetic lines are decoupled. Indeed, the particle dynamics
is controlled by electric field $E=E_\parallel$ which
is related to the instantaneous charge distribution by Gauss'
law (\ref{eq:Gauss1}).
Thus, $E(l)$ on a given magnetic line is fully determined by
charge density $\rho(l)$ {\it on the same line}. The plasma and electric
field evolve along the line as if the world were 1-D.
                                                       
If the conduction current $j$ is smaller than $\jB$,
then $\partial E_\parallel/\partial t>0$ and an electric field appears
that tends to increase the current. Alternatively, if $j>\jB$ then
an electric field of the opposite sign develops which tends to
reduce the current. Thus, $j$ is always regulated toward $j_B$.
This is the standard self-induction effect.
The timescale of the regulation, $\tau$, is very short, of the 
order of the Langmuir oscillation timescale $\omega_{P}^{-1}\sim 10^{-13}$~s. 
Local deviations from charge neutrality tend to be erased on the same
plasma timescale.
The net charge imbalance of the corona must be extremely small ---
a tiny imbalance would create an electric field 
that easily pulls out the missing charges from the surface.
When the corotation charge density is neglected, the net charge of the 
corona vanishes.

Adopting a characteristic velocity $v\sim c$ of the coronal particles,
the Debye length of the plasma $\lD=v/\omega_P$ can be 
taken equal to the plasma skin depth $c/\omega_P$. An important physical
parameter of the corona is the ratio of the Debye length 
to the circuit size $L\sim\RNS$,
\be
\label{eq:zeta}
  \zeta=\frac{\lD}{L}=\frac{c}{L\omega_P}\ll 1.
\ee

The force-free condition $\bj_B\times\bB$ together with $\nabla\cdot\bj_B=0$
requires that $j_B(l)\propto B(l)$ along the magnetic
line.   We can scale
out this variation simply dividing all local quantities (charge density,
current density, and electric field) by $B$. This reduces the problem to an
equivalent problem where $\jB(l)=const$. Furthermore, only
forces along magnetic lines control the plasma dynamics,
and the curvature of magnetic lines falls out from the problem.
Therefore, we can set up the experiment so that plasma
particles move along a straight line connecting anode and cathode (Fig.~2).
We designated this line as the $z$-axis, so that $l=z$.

\begin{figure}
\centering
\includegraphics[width=0.48\textwidth]{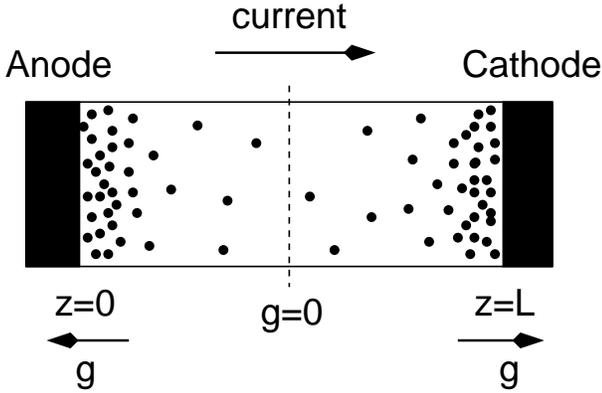}
\caption{
Set up of numerical experiment. Thin and dense plasma layers
are maintained near the cathode and anode by injecting cold particles
through the boundaries of the tube (footpoints of a magnetic flux tube).
The electric current is kept constant at the boundaries and the system
is allowed to evolve in time until a quasi-steady state is reached.
Without voltage between anode and cathode, the current cannot flow
because gravity $g$ traps the particles near the boundaries.
The constant current at the boundaries, however, implies that
voltage is immediately generated should the flow of charge stop
in the tube. The experiment aims to find the self-induction
voltage that keeps the current flowing.
}
\label{fig:1}       
\end{figure}
                                                         
This one-dimensional system may be simulated numerically.
Consider $N$ particles moving along the $z$-axis ($N\sim 10^6$ in our
simulations). The electric field acting on a particle at a given $z$
is given by
\be
\label{eq:Gauss}
   E(z)=4\pi\int_0^z\rho\,\dd z,
\ee
where $E(0)=E(L)=0$ are the boundary conditions.
Given the instantaneous positions $z_i$ of all charges $e_i$ we immediately 
find $E(z)$. To find a quasi-steady state of such a system we let it relax 
by following the particle motion in the self-consistent electric field.
                                                                                
Consider first a circuit where $e^\pm$ production is (artificially) forbidden, 
so that only the cold hydrostatic atmosphere can supply particles to the 
corona.
We find that such a circuit quickly relaxes to a state where it
acts as an ultra-relativistic linear accelerator (Fig.~3). In this state,
oscillations of electric field (on the plasma timescale $\omega_P^{-1}$)
are confined to the thin atmospheric layers. A static accelerating electric
field is created above the layers where the atmosphere density is
exponentially reduced. 
The asymmetry of the solution is caused by the
difference between the electron and ion masses.

\begin{figure}
\centering
\includegraphics[width=0.52\textwidth]{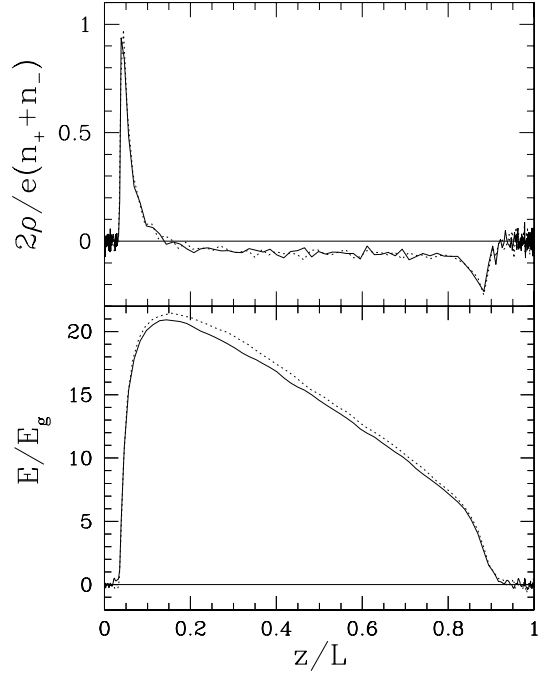}
\caption{
Circuit without $e^\pm$ production. Upper panel shows the normalized
charge density and lower panel shows the electric field in units of
$E_g$ defined by $eE_g=m_ig_0$. Solid and dotted curves correspond to two
different moments of time, demonstrating that a quasi-steady state has
been reached. In this simulation, $m_i=10m_e$ and
$\zeta=c/\omega_PL\simeq 0.01$. The final state is the relativistic
double layer described analytically in \cite{C}.
}
\label{fig:1}       
\end{figure}

This configuration is a relativistic double layer \cite{C}.
It is well described by Carlqvist's solution, which has no gravity
in the circuit and assumes zero temperature at the boundaries, so that
particles are injected with zero velocity. According to this solution,
the potential drop between anode and cathode, in the limit 
$e\Phi_e\gg m_ic^2$, is 
\be
\label{eq:PhiC}
   e\Phi_e = \left[\left(\frac{m_i}{Zm_e}\right)^{1/2}+1\right]
   \,\frac{L\omega_P}{2}\, m_ec, \quad
   \omega_P^2\equiv\frac{4\pi j e}{m_ec},
\ee
where $Z$ is the ion charge number (in our experiment $Z=1$ was
assumed). The established voltage is much larger than is needed to
overcome the gravitational barrier $\Phi_g$. It does not even depend on
$\Phi_g$ as long as $\Phi_g$
is large enough to prohibit the thermally-fed regime.
Gravity causes the transition to the
linear-accelerator state, but the state itself does not depend on $\Phi_g$.

Thus, the first result of numerical experiment is that
an ion-electron circuit (without $e^\pm$ creation) in a
gravitational field relaxes to the double layer of macroscopic size
$L$ and huge voltage $\Phi_e$.
The system does not find any state with a lower $\Phi_e$, even though it
is allowed to be time-dependent and
strong fluctuations persist in the atmospheric layers (reversing the sign
of $E$).

If the linear accelerator were maintained in the twisted magnetosphere,
the twist would be immediately killed off: the huge voltage implies a large
untwisting rate (\S~2). The electron Lorentz factor developed in the 
linear
accelerator is (taking the real $m_i/m_e=1836$ and $\zeta\sim 3\times 10^{-9}$),\be
\label{eq:gamma}
    \gamma_e=\frac{e\Phi_e}{m_ec^2}\simeq \frac{20}{\zeta}
    \sim 6\times 10^9.
\ee
However, new processes will become important before the particles could
acquire such high energies: production of $e^\pm$ pairs will take place.
Therefore, the linear-accelerator
solution cannot describe a real magnetosphere.
We conclude that pair creation is a key ingredient of the circuit
that will regulate the voltage to a smaller value.


\section{Pair discharge and self-organized criticality}

In the magnetospheres of canonical radio pulsars with $B\sim 10^{12}$~G,
$e^\pm$ pairs are created when seed electrons are
accelerated to large Lorentz factors $\gamma_e\sim 10^7$. Such electrons
emit curvature gamma rays that can convert to $e^\pm$ off the magnetic field.
In stronger fields, another channel of $e^\pm$ creation appears.
It is also two-step: an accelerated particle resonantly upscatters a
thermal X-ray photon, which subsequently converts to a pair
\cite{HA01,BT}.
The resonant scattering requires the electron to have a Lorentz factor,
\be
\label{eq:gres}
  \gres = {B/B_{\rm QED}\over1-\cos\theta_{kB}}\,
            \frac{m_ec^2}{\hbar\omega_X}
      \approx 10^3\,B_{\rm 15}\,\frac{\rm 10~keV}{\hbar\omega_X},
\ee
where $\hbar\omega_X$ and $\theta_{kB}$ are the energy and pitch angle
of the target photon.
The distance over which an accelerated electron creates a pair is
approximately equal to the free path to scattering $\lres$.
It depends only on $B$ and the spectrum of target photons;
$\lres\ll\RNS$ for the relevant parameter $B_{15}\hbar\omega_X<100$~keV
\cite{BT}.

When the voltage in the circuit becomes high enough to accelerate electrons
to $\gres$, an $e^\pm$ breakdown develops in the magnetic flux tube.
An illustrative toy model of breakdown is shown in a spacetime diagram 
in Figure~4. 
Each pair-creation event gives two new particles of opposite charge,
which initially move in the same direction. One of them is accelerated by
the electric field and lost after a time $<L/c$,
while the other is decelerated and can reverse direction before reaching
the boundary.  This reversal of particles in the tube 
and repeated pair creation
allows the $e^\pm$ plasma to be continually replenished.
In the super-critical regime
(left panel in Fig.~4) more than one reversing
particle is created per passage time $L/c$ and an avalanche develops
exponentially on a timescale $\sim \Lres/c$.
In the near-critical regime shown in the right panel,
just one reversing particle is created per passage time $L/c$.
This critical state is unstable: sooner or later the avalanche will
be extinguished.

\begin{figure}
\centering
\includegraphics[width=0.52\textwidth]{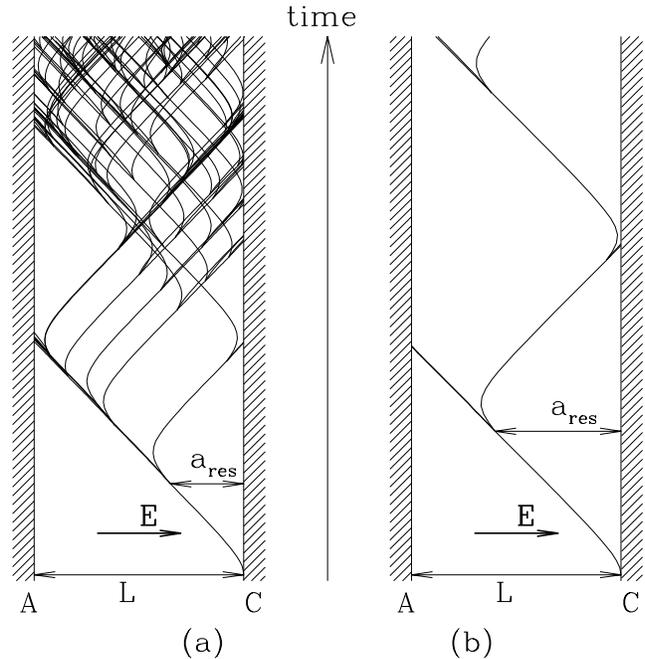}
\caption{
Spacetime diagrams illustrating the critical character of the
$e^\pm$ breakdown and formation of avalanches. The seed electron is placed 
at cathode and is accelerated to $\gres$ after passing the distance 
$\Lres=(\gres-1)m_ec^2/eE$.
The figure shows the worldlines of created particles.
(a) $\Lres/L=0.35$ (supercritical case).
(b) $\Lres/L=0.6$ (critical case).
}
\label{fig:4}       
\end{figure}

This toy model shows an essential property of the $e^\pm$
breakdown: it is a critical stochastic phenomenon.
Above a critical voltage pair creation proceeds in a runway manner,
and the current and the dissipation rate would run away if the voltage
were fixed. Below the critical voltage pair creation does not ignite.
The criticality
parameter is $L/\Lres=e\Phi_e/(\gres-1)m_ec^2$. The tube with enforced
current at the boundaries must self-organize to create pairs in the
near-critical regime $e\Phi_e\sim \gres m_ec^2$ and maintain the current.
                                                                                
The discharging tube is similar to other phenomena that show self-organized
criticality, e.g., a pile of sand on a table \cite{Bak}.
If sand is steadily added, a quasi-steady state is established with a
characteristic mean slope of the pile. The sand is lost (falls from the
table) intermittently, through avalanches --- a sort of ``sand discharge.''
In our case, charges of the opposite signs are added steadily instead of
sand (fixed $j$ at the boundaries), and voltage $\Phi_e=EL$ plays the role
of the mean slope of a pile. The behavior of the discharging system is
expected to be time-dependent, with stochastic avalanches.

We implemented the process of pair creation in our numerical experiment.
Regardless the details of this process (values of $\gres$ and $\lres$) 
the circuit relaxed to the state of self-organized criticality 
after a few $\tdyn=L/c$.
This state is time-dependent on short timescales, but it has a
well-defined steady voltage when averaged over a few $\tdyn$ (Fig.~5),
\be
 e\bar{\Phi}_e\sim \gres m_ec^2.
\ee
  We found that 
this relation applies even to circuits with $\gres m_ec^2\gg m_ic^2$,
where lifting of the ions is energetically preferable to pair creation.
We conclude that {\it the voltage along the magnetic tube is self-regulated
to a value just enough to maintain pair production and feed the current
with $e^\pm$ pairs}. In all cases, 
the pair creation rate $2\dot{N}_+\sim \jB/e$ is maintained.

The solution for the plasma dynamics in the corona is strongly
non-linear and time-dependent. It is essentially global in the sense that
the plasma behavior near one footpoint of a magnetospheric field line is
coupled to the behavior near the other footpoint. Remarkably, this
complicated global behavior may be described as essentially one-dimensional
electric circuit that is subject to simple boundary conditions $E\approx 0$
on the surface of the star and can be studied using a direct numerical
experiment.
                                                                                
\begin{figure}
\centering
\includegraphics[width=0.52\textwidth]{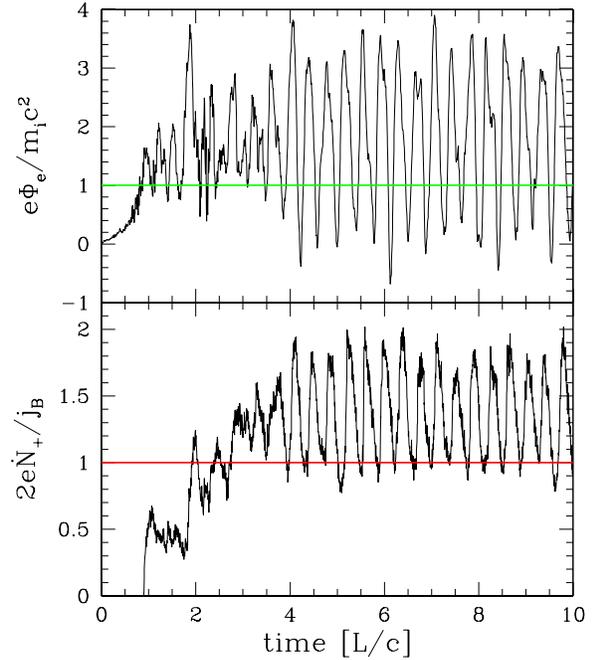}
\caption{ Time history of one numerical experiment.
The circuit parameters are: $m_i=10m_e$, $\gres=m_i/m_e$, $\zeta=0.01$. 
After a few dynamical times the voltage $\Phi_e$ stops growing and the 
circuit enters an oscillatory regime. During each oscillation, an increased 
voltage triggers $e^\pm$ discharge, then 
$\dot{N}_+$ drops, and $\Phi_e$ begins to grow again.
The quasi-steady oscillating state (self-organized criticality) is 
established. 
The horizontal lines show the voltage $e\Phi_e=\gres m_ec^2$ and the minimum 
pair production rate $2\dot{N}_+=\jB/e$ that can feed the required current 
$\jB$.
}
\label{fig:5}       
\end{figure}

The current is carried largely by $e^\pm$ everywhere in the tube.
The ion fraction in the current depends on the ratio
$e\Phi_e/m_ic^2\simeq\gres m_e/m_i$ (Fig.~6). If this ratio is small,
pairs are easily produced with a small electric field, too small to lift
the ions, and the ion current is suppressed. In the opposite case, ions
carry about 1/2 of the current. By coincidence, $\gres\simlt m_i/m_e$ 
is typical for magnetars, which implies that ions carry $\simlt 10$\% 
of the current.

\begin{figure}
\centering
\includegraphics[width=0.49\textwidth]{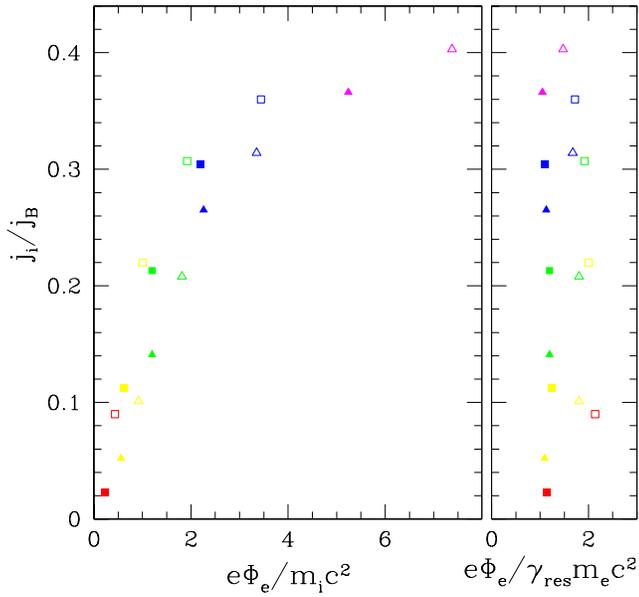}
\caption{
Fraction of the electric current carried by ions vs. voltage.
In the left panel voltage is taken in units of $m_ic^2$,
and in the right panel --- in units of $\gres m_ec^2$.
Different colors and symbols show circuits with $m_i/m_e=10,30$ and 
various values of $\gres$ ranging from $m_i/5m_e$ to $5m_i/m_e$ 
(see \cite{BT}). 
}
\label{fig:6}       
\end{figure}

\section{Implications}

The basic finding of this work is that an $e^\pm$ corona must be
maintained around a magnetar. It exists in a state of self-organized
criticality, near the threshold for $e^\pm$ breakdown. The stochastic
$e^\pm$ discharge continually replenishes the coronal plasma, which
is lost on the dynamic time $t_{\rm dyn}\sim 10^{-4}$~s.
The established voltage along a twisted magnetic tube is
marginally sufficient to accelerate an electron (or positron) to the
energy where collisions with ambient X-ray photons spawn new pairs.
This physical condition determines the rate of energy dissipation in
the corona. The generation of this voltage is precisely the
self-induction effect of the gradually decaying magnetic twist, and
the energy release in the corona is fed by the magnetic energy
of the twisted field.

The rate of energy dissipation in the twisted magnetosphere is given by
$L_{\rm diss}=I\Phi_e$ where $I$ is the net current through the corona.
The current may be estimated as
$I\sim \jB a^2\sim \frac{c}{4\pi}\,\Delta\phi\,(B/\RNS)\,a^2$,
where $a$ is the size of a twisted region on the stellar surface and
$\Delta\phi=|\nabla\times\bB|(B/\RNS)^{-1}\simlt 1$ characterizes the
strength of the twist. The calculated $e\Phi_e\sim 1$~GeV implies
\be
\label{eq:Ldiss}
   L_{\rm diss}=I\Phi_e\sim 10^{37}\,\Delta\phi\,B_{15}
    \left(\frac{a}{\RNS}\right)^2\left(\frac{e\Phi_e}{\rm GeV}\right)
           \,\frac{\rm erg}{\rm s}.
\ee
Observed luminosity $\Ldiss \sim 10^{36}$~erg~s$^{-1}$
is consistent with a partially twisted magnetosphere, $a\sim 0.3\RNS$, or
a global moderate twist with $\Delta\phi\,(e\Phi_{e}/\rm GeV)\sim 0.1$.
                                                                                
Once created, a magnetospheric twist has a relatively long but finite
lifetime. The energy stored in it is
${\cal E}_{\rm twist}\sim(I^2/c^2)\RNS$.  This energy dissipates in a time
\begin{eqnarray}
\label{eq:tdecay}
     t_{\rm decay}=\frac{{\cal E}_{\rm twist}}{\Ldiss}\approx
        3\,\left(\frac{L_{\rm diss}}{10^{36}\rm ~erg~s^{-1}}\right)
                \left(\frac{e\Phi_e}{\rm GeV}\right)^{-2} {\rm ~yr}.
\end{eqnarray}
Note that a stronger twist (brighter corona) lives longer. If the time 
between large-scale starquakes is longer than $t_{\rm decay}$,
the magnetar should be seen to enter a quiescent state.
Such behavior has been observed in AXP~J1810-197: an outburst
was followed by the gradual decay on a year timescale \cite{GH05}.
                                                                                
The voltage $\Phi_e$ implies a certain effective resistivity
of the corona, ${\cal R}=\Phi_e/I$. This resistivity leads to
spreading of the electric current {\it across} the magnetic lines, which
is described by the induction equation 
$\partial\bB/\partial t=-c\nabla\times\bE$.  
The timescale of twist spreading to the light cylinder is comparable 
to $t_{\rm decay}$.
Therefore, the impact of a starquake on spindown may appear with a 
delay of $\sim $ years. 
                                                                                
Even a small ion current implies a large mass transfer through the
magnetosphere over the $\sim 10^4$-yr active lifetime of a magnetar.
It requires a significant excavation of the crust at the anode footpoints 
of twisted magnetic tubes.
The bombarding relativistic electrons from the corona cause this excavation.
They spall heavy ions in the uppermost crust, re-generate the
light-element atmosphere on the surface, and regulate its column density
to a value $\sim 100 m_p/\sT$ \cite{BT}.

The radiative output from the corona likely peaks in its inner region because
the coronal current is concentrated on closed field lines with a 
maximum extension $\Rmax\simlt 2\RNS$ (\S~2). 
Emission from the inner corona must be suppressed above $\sim 1$~MeV,
regardless the mechanism of emission, because photons with energy
$\simgt 1$~MeV cannot escape the ultra-strong magnetic field. 
The observed high-energy radiation extends above 100~keV 
\cite{Kuiper06,Kuiper04,Mer05,Mol05,denH06}.
There is an indication for a cutoff between 200~keV and 1~MeV from
COMPTEL upper limits for AXP 4U 0142$+$61.
                                                                                
The possible mechanisms of emission are strongly constrained.
The corona has a low density
$n\sim n_c=\jB/ec\simlt B(4\pi\,e\RNS)^{-1}\sim 10^{17}B_{15}$~cm$^{-3}$
and particle collisions are rare, so two-body
radiative processes are negligible. 
The upscattering of keV photons streaming from the surface does not
explain the observed 100-keV emission \cite{BT}. A possible source of 
the 100~keV X-rays is the transition
layer between the corona and the thermal photosphere \cite{TB05,BT}.
It is heated by the coronal beam and cooled by heat conduction toward the 
surface, and its temperature is regulated to
 $\sim 100(\ell/\lCoul)^{-2/5}$~keV, where $\ell/\lCoul<1$
parameterizes the suppression of thermal conductivity by plasma turbulence.
The emitted X-ray spectrum may be approximated as single-temperature 
optically thin bremsstrahlung \cite{TB05}.
Its photon index below the exponential cutoff is close to $-1$, in agreement 
with spectra observed with INTEGRAL and RXTE.
                                                                                
The observed pulsed fraction increases toward the high-energy end of the
spectrum and approaches 100\% at 100~keV (Kuiper et al. 2004).
If the hard X-rays are
produced by one or two twisted spots on the star surface, the large
pulsed fraction implies that the spots almost disappear during some phase
of rotation. Then they should not be too far from each other.
They cannot be, e.g., antipodal because a large fraction of the
star surface is visible to observer due to the gravitational bending
of light: $S_{\rm vis}/4\pi\RNS^2=(2-4GM/c^2\RNS)^{-1}\approx 3/4$
\cite{B}.

The transition layer at a footprint of a coronal flux tube may reach 
high enough temperature for thermal $e^\pm$ production \cite{BT}. The created 
pairs then elevate above the surface and can conduct the magnetospheric 
current. This opens a possibility for the corona to be fed by surface pair 
creation instead of the $e^\pm$ discharge.
                                                                                
The observed infrared ($K$-band) and optical luminosities of magnetars are
$\sim 10^{32}$~erg/s \cite{Hul04},
which is far above the
Rayleigh-Jeans tail of the surface blackbody radiation.  The
inferred brightness temperature is $\sim 10^{13}$~K if the radiation
is emitted from the surface of the star. Four
principle emission mechanisms may then be considered: coherent plasma
emission, synchrotron emission by electrons, cyclotron emission by ions,
and curvature radiation by $e^\pm$ bunches in the corona. The last two 
are most plausible (see \cite{BT} for detailed discussion). In particular,
curvature radiation in the inner magnetosphere may extend into the 
optical-IR range.  Charges moving on field lines with
a curvature radius $R_C \simlt R_{\rm NS}$ will emit
radiation with frequency $\nu_C\sim\gamma_e^3 c/2\pi R_C$,
which is in the K-band ($\nu = 10^{14}$~Hz) or optical when
$\gamma_e m_ec^2\simgt$~GeV.
The radio luminosity of a normal pulsar can approach $\sim 10^{-2}$ of
its spindown power, and it is not implausible that the observed fraction
$\sim 10^{-4}$ of the power dissipated in a magnetar corona
is radiated in optical-IR photons by the same coherent mechanism.

The corona luminosity is simply proportional to the net current 
flowing through the magnetosphere, and is determined by the competition
between occasional twisting (due to starquakes) and gradual dissipative
untwisting of the magnetosphere. During periods of high activity, when 
starquakes occur frequently, the magnetospheric twist
may grow to the point of a global instability:
when a critical $\Delta\phi\simgt 1$ is achieved,
the magnetosphere suddenly relaxes to a smaller twist angle. 
A huge release of energy must then occur, producing a giant flare;
a model of how this can happen is discussed in \cite{L06}.
                                                                                
A possible way to probe the twist evolution is to measure the history
of spindown rate $\dot{P}$ of the magnetar. The existing data confirm
the theoretical expectations: the spin-down of SGRs was observed to 
accelerate months to years following periods of activity \cite{Kouv99,Woods02}.
A similar effect was also observed following 
flux enhancement in AXPs \cite{GK04}.
This ``hysteresis'' behavior of the spindown rate may represent the delay 
with which the twist is spreading to the outer magnetosphere.
                                                                                
Since the density of the plasma corona is proportional to the current that
flows through it, changes in the twist
also lead to changes in the coronal opacity.
The density of the plasma corona is close to its minimum $n_c=j/ec$,
and it is largely made of $e^\pm$ pairs rather than electron-ion plasma. 
The opacity should increase following bursts of activity and affect
the X-ray pulse profile and spectrum.
Such changes, which persist for months to years, have been observed
following X-ray outbursts from two SGRs \cite{Woods01}.

\begin{acknowledgements}
This work was supported by NASA grant NNG-06-G107G.
\end{acknowledgements}



\end{document}